# A Tandem Cell for Nanopore-based DNA Sequencing with Exonuclease

G. Sampath
sampath_2068@yahoo.com

**Abstract**
A tandem cell is proposed for DNA sequencing in which an exonuclease enzyme cleaves bases (mononucleotides) from a strand of DNA for identification inside a nanopore. It has two nanopores and three compartments with the structure [*cis*1, upstream nanopore (UNP), *trans*1=*cis*2, downstream nanopore (DNP), *trans*2]. The exonuclease is attached to the exit side of UNP in *trans*1/*cis*2. A cleaved base cannot regress into *cis*1 because of the remaining DNA strand in UNP. A profiled electric field over DNP with positive and negative components slows down base translocation through DNP. The proposed structure is modeled with a Fokker-Planck equation and a piecewise solution presented. Results from the model indicate that with probability approaching 1 bases enter DNP in their natural order, are detected without any loss, and do not regress into DNP after progressing into *trans*2. Sequencing efficiency with a tandem cell would then be determined solely by the level of discrimination among the base types inside DNP.

## I INTRODUCTION

The success of the Human Genome Project has seen increased efforts aimed at efficiently sequencing whole genomes. A variety of methods have been developed for high throughput sequencing of DNA molecules of varying lengths based on a wide range of technologies that may be label-based or label-free. Most of them require considerable amounts of preprocessing, preparation, and reagents. Comprehensive information about these methods can be found in several reviews [1, 2].

Nanopore-based methods use a minimal amount of preparation and rely primarily on the detection of a current in the pore when the analyte (usually a DNA molecule or fragment thereof, but more generally any polymer) passes through. Such sequencing is efficient in principle and leads to a device footprint that is smaller than in the other NGS methods. Nanopore sequencing has been reviewed in detail recently [3, 4].

The present work proposes a nanopore-based sequencing structure that has two nanopores in tandem and an exonuclease enzyme that is attached to the output side of the first nanopore and cleaves bases from an ssDNA molecule that has threaded through the first pore. The cleaved bases pass through the second pore and are detected there. A profiled voltage applied across the second pore slows down translocation of the base thus reducing the detection bandwidth required. The structure is modeled with a Fokker-Planck equation and a piecewise solution presented. Using results from recent experiments with exonuclease-based sequencing [5] in which single bases are found to cause blockade levels unique to a base type, the physical model shows that in theory a tandem cell is capable of efficiently sequencing a single-stranded DNA molecule.

## II NANOPORE-BASED SEQUENCING OF DNA

Figure 1a shows the basic structure of a conventional nanopore-based electrolytic cell. Negative (positive) ions flow through the pore from *cis* (*trans*) to *trans* (*cis*) under the influence of a positive (negative) potential difference. This structure can be used to identify the sequence of bases in a strand of DNA.

In 'strand sequencing' (Figure 1a) a strand of DNA is introduced into *cis*, and the negatively charged biomolecule is drawn into the pore. Successive bases (A, T, C, G) passing through the pore are identified by the current blockade level [6]. The level of discrimination depends on the length and width of the pore and the speed with which the strand translocates through the pore.

In 'exonuclease sequencing' (Figure 1b), an exonuclease enzyme in *cis* next to the vestibule of an AHL nanopore cleaves single bases in ss-DNA and drops them into the pore where they are identified by the current blockade level [5]. This method has a number of problems such as cleaved bases diffusing back into the *cis* compartment to be 'lost' or called out of order, two or more cleaved bases occupying the pore at the same time, and excessive translocation speeds. The method has been modeled mathematically and analyzed in detail [7].

Pores may be biological or synthetic. Biological pores include AHL [8] and MspA [9]. Synthetic pores may use silicon nitride ($Si_3N_4$) and/or graphene [10-12]. In the latter case changes in the lateral current flowing through the sub-nanometer thick graphene are used to uniquely identify bases when ssDNA translocates from to *cis* to *trans* through a nanopore in the graphene [13]. Other synthetic pores studied include 'DNA transistors' [14] and silicon-based gated MOSFET-like structures [15, 16].

In the following sections a tandem nanopore structure based on the exonuclease-based DNA sequencing method described above [5] is proposed and analyzed theoretically.

**III TANDEM CELL: PHYSICAL STRUCTURE AND RATIONALE**
A tandem cell (Figure 2) consists of two conventional electrolytic nanopore-based cells connected in tandem and an exonuclease enzyme in between. It has MspA or AHL for the first (upstream) pore (UNP) with the exonuclease enzyme covalently bonded to it on the *trans* side, a silicon microchannel for *trans*1/*cis*2, and a solid-state pore for the second (downstream) pore (DNP). A potential difference is applied through electrodes situated at the top of *cis*1 and the bottom of *trans*2. The enzyme cleaves bases from a single-stranded DNA (ssDNA) molecule that has translocated through UNP, and the cleaved bases are detected during their passage through DNP.

Formally the tandem cell can be written as [*cis*1, UNP, *trans*1=*cis*2, DNP, *trans*2], a pipeline of five sections numbered 1 through 5, with 0 referring to the top of *cis*1. The following notation is used: $V_{ij}$ is the potential difference between sections i and j, $L_{i-1\ i}$ the length of section i, $E_{i-1\ i}$ the electric field across it, and $v_{i-1\ i}$ the drift velocity. $V_i$ is the applied voltage at the bottom of section i (or the top of section i+1) and also represents a physical electrode. $V_0$ is the voltage at the top of *cis*1, asssumed without loss of generality to be held at ground. Minor variations are used as convenient.

**Rationale**
The tandem cell requires a single DNA strand to be captured in the mouth of UNP, thread through UNP, and present itself to the exonuclease enzyme on the *trans* side of UNP (*trans*1/*cis*2) to be cleaved. Capture-threading has been experimentally found to always occur if the voltage across the pore is sufficiently large, see Figure 7 in [3]. This can be realized in practice with the tandem cell for the following reason. The bulk of the applied voltage across the *cis* and *trans* compartments of a conventional cell drops across the pore, with only about 1% over the *cis* and *trans* compartments [5, 7]. In a tandem cell, for an applied voltage $V_{05}$ = 0.4 V assume that 49.5% of the voltage drops across each of the two pores. This means ~198 mV across UNP, which is sufficient to ensure capture-threading of the DNA in and through UNP. For this reason in the model below it is assumed that ssDNA is always captured in the mouth of UNP, threads through UNP to the top of *trans*1/*cis*2 for cleaving by the enzyme, and is delivered to DNP for detection.

For sequencing to be accurate it is necessary (in a statistical sense) that: a) cleaved bases arrive at and be captured by DNP in their natural order; b) DNP identify each and every base as it passes through; and c) the detected base exit DNP without regressing.

All these conditions are satisfied in the tandem cell. This statement is justified informally below, with a more formal quantitative analysis provided in Section IV.

1) The leading base of the ssDNA is cleaved by the enzyme on the *trans*1/cis2 side of UNP at a rate that varies from one every 10 msecs [5, 7] to one every 80 msecs [17]. Thus on the average cleaved bases are separated in time by at least 10 msecs. Now consider a cleaved base drifting toward DNP under the influence of $V_{05}$. With 99% of the applied voltage dropping across the two pores, the remaining 1% of the applied voltage of 0.4 V drops across *cis*1, *trans*1/*cis*2, and *trans*2. Assuming for simplicity that it is divided among them in the proportion 3:4:3 (the three compartments are roughly of the same height and contain the same electrolyte), there is a drop of 1.6 mV across *trans*1/*cis*2. With a length of 1 μm and mobility μ = 2.4 × $10^{-8}$ m/volt-sec, the electric field is 1600 V/m and the drift velocity is 3.84 × $10^{-5}$ m/sec. The mean translocation time for a cleaved base to drift-diffuse through *trans*1/*cis*2 is then about 25 msecs. If the spread (standard deviation) of the translocation time is not too high then successive bases (which are cleaved at the rate of 1 every 10 msecs) will not enter DNP out of order. Also a cleaved base cannot regress into *cis*1 because the remaining DNA strand blocks its passage.

2) With ~198 mV dropping across DNP, DNP length of 10 nm, and base mobility μ = 2.4 × $10^{-8}$ $m^2$/volt-sec (assumed to be the same for all base types), the mean translocation time through DNP is about 20 nsecs, which is much less than the time between two successive bases arriving at DNP (~10 msecs). Thus two bases do not occupy DNP at the same time.

3) A detected base exits into *trans*2 under the influence of $V_{05}$. With 198 mV across DNP (which is close to the the optimum value of 180 mV given in [5]) an exiting base will not regress into the pore from *trans*2. The likelihood of regression can be further reduced by providing a reinforcing drift field across *trans*2. Thus two electrodes may be used in *trans*2 one at the top and the other at the bottom and a voltage difference applied across them (Figure 3). (In this case the height of *trans*2 can be larger than 1 μm.)

From 1) through 3) one can conclude that with a tandem cell bases enter DNP in the correct order (with no more than one occupying DNP at the same time), are not skipped, and having passed through do not regress into DNP. Assuming that DNP correctly identifies each base (that is, blockade levels are sufficiently discriminating of base type), the tandem cell can effectively sequence a single strand of DNA.

## IV ANALYTICAL MODEL

The behavior of a cleaved base as it translocates through *trans*1/*cis*2 and DNP can be studied via the trajectory of a particle whose propagator function G (x,y,z,t) is given by a linear Fokker-Planck (F-P) equation in three dimensions. Such methods are commonly used in the study of translocation of biomolecules through a nanopore [7]. The F-P equation is used for piecewise analysis of the propagator in two sections: *trans*1/*cis*2 and DNP. Each section is modeled independently in its own coordinate system and the transition occurring at the interface between the two stages is studied separately. The coordinate systems used are shown in Figure 4. Standard methods from partial differential equations and Laplace transforms are used [18, 19].

**Translocation through DNP** (*Detection*)
A one-dimensional approximation is applied to DNP (Figure 4a). The trajectory of the cleaved base as it passes through DNP is described by the function G(z, t) which satisfies

$$\partial G/\partial t + v_z \partial G/\partial z = D \partial^2 G/\partial z^2 \qquad z \in [0, L=L_{34}] \qquad (1)$$

with initial and boundary value conditions

I.V. The particle is released at z = 0 at time t = 0:

$$G(0, t = 0) = \delta(z) \qquad (2)$$

B.C.1 The particle is captured at z = L:

$$G(L, t) = 0 \qquad (3)$$

B.C.2 The particle is reflected at z = L:

$$D \partial G(z, t)/\partial z |_{z=0} = v_z G(z, t) \qquad (4)$$

Here $v_z$, the drift velocity through DNP, is given by $v_z = \mu V_{34}/L$, and $\mu$ is the nucleotide mobility (assumed to be the same for all four types). Following standard procedures φ(t), the pdf of the first passage time (translocation time) for a particle to diffuse-drift from z = 0 to z = L and get absorbed at z = L, can be obtained as

$$\varphi(t) = (2/(\sqrt{\pi 4Dt^3}) [ \sum_{k=0}^{\infty} ((2k+1)L + v_z t) \exp(-((2k+1)L + v_z t)^2/(4Dt)) + \sum_{k=0}^{\infty} ((2k+1)L - v_z t) \exp(-((2k+1)L + v_z t)^2/(4Dt)) \qquad (5)$$

φ(t) can be computed numerically but the series oscillates and converges very slowly. Therefore an alternative closed-form approach based on the earlier referenced model [7] of exonuclease-based sequencing. In that model a base is assumed to be cleaved above the pore of a conventional cell and drop into the pore. This results in a non-zero probability of a cleaved base not entering the pore (given by a rate constant κ) and getting lost to diffusion. By setting κ to 0 the model is reduced to the boundary value problem in Equations 1-4. Also unlike in [7] where the drift velocity is assumed to be always oriented in the downstream direction (*cis* to *trans*), here it is assumed that the drift velocity $v_z$ can be positive or negative.

Of interest here is φ(t), the pdf of the first passage time T (the time for a cleaved base to translocate through the pore and be detected at the end of its translocation), which is independent of the coordinate system. Modifying the main result in [7] the Laplace transform of the first passage time of a cleaved base passing through DNP is

$$\varphi^*(s) = \exp(\alpha/2) / [\cosh(y) + \alpha/2 \ \sinh(y)/y] \qquad (6)$$

where

$$\alpha = v_z L/D; \qquad y^2 = \sqrt{(\alpha^2/4 + 2\tau s)}; \qquad \tau = L^2/2D \qquad (7)$$

The mean $E(T)$ is

$$E(T) = -d\varphi^*(s)/ds \, |_{s=0} = (L^2/D\alpha)[1 - (1/\alpha)(1 - \exp(-\alpha))] \qquad (8)$$

Similarly, the second moment $E(T^2)$ can be obtained as:

$$E(T^2) = d^2\varphi^*(s)/ds^2 \, |_{s=0} = 2(L^2/D\alpha^2)^2(\alpha^2/2 + 3\alpha \exp(-\alpha) - 2 + \exp(-\alpha) + \exp(-2\alpha)) \qquad (9)$$

From here the variance $\sigma^2(T) = E(T^2) - E^2(T)$ is obtained as

$$\sigma^2(T) = (L^2/D\alpha^2)^2 (2\alpha + 4\alpha \exp(-\alpha) - 5 + 4 \exp(-\alpha) + \exp(-2\alpha)) \qquad (10)$$

where $\sigma$ is the standard deviation.
For $v_z = 0$, these three statistics are given by

$$E_0(T) = L^2/2D \qquad E_0(T^2) = (5/12)(L^4/D^2) \qquad \sigma_0^2(T) = (1/6)(L^4/D^2) \qquad (11)$$

Figure 5 shows the mean and standard deviation of T for different voltages across a DNP with L = 10 nm, D = 3 × 10$^{-10}$ m$^2$/sec, and μ = 2.4 × 10$^{-8}$ m$^2$/volt-sec. In [5] the optimum potential difference across the nanopore for detecting a base dropped into the pore of a conventional cell is noted as 0.180 V. For this value the mean and standard deviation of the translocation time are on the order of 10$^{-8}$ sec. The resulting bandwidth is very high, on the order of 40 MHz. (Figure 5 shows translocation statistics for both positive and negative values of the voltage across DNP. Later in this section the possibility of slowing down the translocating base and thereby reducing the bandwidth requirement using a negative voltage across part of DNP is considered.)

**Translocation through** *trans*1/*cis*2 (*Delivery*)
For simplicity the *trans*1/*cis*2 compartment is assumed to be a rectangular box-shaped region (Figure 4b). A particle is released at the top and translocates to the bottom of the compartment where it is 'absorbed'. 'Absorption' here means that the particle moves into DNP without regressing into *trans*1/*cis*2. Its behavior in DNP is described by the model pertaining to that section. The propagator function G(x, y, z, t) is given by a linear Fokker-Planck equation in three dimensions:

$$\partial G/\partial t + v_x \, \partial G/\partial x + v_y \, \partial G/\partial y + v_z \, \partial G/\partial z = D \, (\partial^2 G/\partial x^2 + \partial^2 G/\partial y^2 + \partial^2 G/\partial z^2) \qquad (12)$$

where $v_x$, $v_y$, and $v_z$ are the drift velocities in the x, y, and z directions, and D is the diffusion coefficient. In *trans1/cis2* there is no drift potential in the x and y directions (Figure 4b) so

$$v_x = v_y = 0 \qquad (13)$$

in Equation 12.

The following initial value (I.V.) and boundary values (B.C.) apply:
1) The particle is released at position (0, 0, 0) at time t = 0. This is represented by a delta function $\delta(x,y,z)$:

I.V. $\qquad G(0, 0, 0, t=0) = \delta(x,y,z) = \delta(x)\,\delta(y)\,\delta(z) \qquad (14)$

2) It is absorbed at the bottom of *trans1/cis2* at t > 0:

B.C. 1 $\qquad G(x, y, L_{23}=L, t) = 0 \qquad (15)$

3) It is reflected at the sides of *trans1/cis2* at t > 0:

B.C. 2 $\qquad D\,\partial G(x, y, z, t)/\partial x\,|_{x=\pm d/2} = 0 \qquad (16)$

B.C. 3 $\qquad D\,\partial G(x, y, z, t)/\partial y\,|_{y=\pm d/2} = 0 \qquad (17)$

3) It is reflected at the top of trans1/cis2:

B.C. 4 $\qquad D\,\partial G(x, y, z, t)/\partial z\,|_{z=0} = v_z\,G(x, y, 0, t), \qquad t > 0 \qquad (18)$

Since the initial value is a separable function of x, y, and z (Equation 14), the above boundary value problem in three dimensions can be considered mathematically as three boundary value problems [18], one in each dimension, and the propagator function viewed as the product of three independent propagator functions:

$$G(x,y,z,t) = G_x(x,t)\,G_y(y,t)\,G_z(z,t) \qquad (19)$$

where

$$G_x(x,t) = (2/d) \sum_{m=0}^{\infty} \cos \alpha_m x/\sqrt{D}\, \exp(-\alpha_m^2 t) \qquad (20)$$

$$G_y(y,t) = (2/d) \sum_{n=0}^{\infty} \cos \beta_n x/\sqrt{D}\, \exp(-\beta_n^2 t) \qquad (21)$$

and

$$G_z(z,t) = (2D/L)\, \exp(v_z z/2D + v_z^2/4Dt) \sum_{k=1}^{\infty} \sin \omega_k L\, \sin \omega_k(z-L)\, \exp(-D\omega_m^2 t)/N(\omega_k) \qquad (22)$$

with

$$\alpha_m = 2m\pi\sqrt{D}/d \qquad \beta_n = 2n\pi\sqrt{D}/d \qquad (23)$$

and

$$N(\omega_k) = (D/v_z)(\exp(v_z L/D)-1) - \{(v_z/D)(\exp(v_z L/D) - \cos 2\omega_k L) - 2\omega_k \sin 2\omega_k L\}/((v_z/D)^2 + 4\omega_m^2) \qquad (24)$$

If detection is defined to occur when the particle reaches $z = L$, the first passage time is the time the particle crosses $z=L$ at any x and y, $-d/2 \leq x,y \leq d/2$, so that its pdf $\varphi(t)$ can be written as

$$\varphi(t) = \int_{-d/2}^{d/2} \int_{-d/2}^{d/2} (-D\, dG(x,y,z,t)/dz\,|_{z=L})\, dx\, dy = \int_{-d/2}^{d/2} G_x(x,t)\, dx \int_{-d/2}^{d/2} G_y(y,t)\, dy\, \varphi_z(t) \quad (25)$$

where

$$\varphi_z(t) = 2D \exp(v_z z/2D - v_z^2/4Dt) \sum_{k=1}^{\infty} \omega_k \sin \omega_k L \exp(-D\omega_m^2 t)/N(\omega_k) \quad (26)$$

Similar to separation of the three-dimensional boundary value problem defined by Equations 12-18 into three independent one-dimensional boundary value problems [18], one can consider in physical terms a similar separation of diffusive effects in the three directions. With free diffusion given by Equations 12-13 and only the initial condition in Equation 14, the diffusion has a spatial mean of (0, 0, 0) and is independent in the three directions. Adding the reflective boundaries $z = 0$, $x = \pm d/2$, and $y = \pm d/2$ (see Figure 2) and a positive drift potential ($V_{23} > 0$) causes the mean of the first passage time to $z = L$ (which is an absorbing boundary, where detection is considered to occur for any x and y; $-d/2 \leq x,y \leq d/2$) to be less than the mean time when $V_{23} = 0$. Considering $\varphi_z(t)$ in isolation, its distribution is in effect the one-dimensional first passage time distribution with mean $E(T = T_z)$ and standard deviation $\sigma = \sigma_z$.

To see if diffusion in the x and y directions has any effect on $G(x,y,z,t)$ consider the factor $\int_{-d/2}^{d/2} G_x(x,t)\, dx$ in Equation 25 (the behavior of $G_y(y,t)$ is identical owing to the symmetry in x and y). To compute it the method of images [19] can be used. Thus start without any boundary conditions on x, which corresponds to free diffusion in x. $G_x(x,t)$ is then given by the heat kernel:

$$G_x(x,t) = (1/(\sqrt{\pi 4Dt})) \exp(-x^2/(4Dt)), \qquad -\infty < x < \infty \quad (27)$$

This function is repeatedly reflected at $x = \pm d/2$ resulting finally in

$$G_x(x, t) = (1/\sqrt{\pi 4Dt})[\exp(-x^2/4Dt) + \sum_{k=1}^{\infty} \exp(-(x+kd)^2/4Dt) + \sum_{k=1}^{\infty} \exp(-(x-kd)^2/4Dt)],$$
$$-d/2 \leq x \leq d/2 \quad (28)$$

Because probability is conserved, the integral of $G_x(x, t)$ over $-d/2 \leq x \leq d/2$ is the area under the heat kernel function over $-\infty < x < \infty$, which is 1. A similar result holds for $G(y, t)$ by symmetry. Hence Equation (25) reduces to

$$\varphi(t) = \varphi_z(t) \quad (29)$$

Thus diffusion in the x and y dimensions does not affect the translocation time in the z dimension, assuming that arrival of the particle at any (x, y, z=L) is tantamount to detection. Figure 6 shows the dependence on pore voltage of the mean $E(T)$ and standard deviation $\sigma(T)$ of the translocation time through *trans*1/*cis*2 for $L = 1$ μm.

**Behavior at the interface between two sections**

Consider the probability currents at some fixed point $(x,y,L_{23}\pm)$ on either side of the interface *trans*1/*cis*2-DNP. If there is an absorbing barrier at $L_{23}$- then the probability function on the *trans*1/*cis*2 side would be

$$G_3(x,y,L_{23}\text{-},t) = 0 \tag{30}$$

On the DNP side if there is a reflecting barrier the probability current would be

$$J_4(x,y,L_{23}\text{+},t) = v_{z4}\, G_4(x,y, L_{23}\text{+},t) - D\, \partial G_4(x,y,L_{23}\text{+},t)/\partial z = 0 \tag{31}$$

$$v_{z3} = \mu V_{23}/L_{23} \qquad v_{z4} = \mu V_{34}/L_{34} \tag{32}$$

But there is really no barrier. The particle oscillates at the interface because of diffusion, before eventually passing into DNP, such passage being aided directly by the positively directed drift potentials in both compartments (and indirectly by the reflecting boundaries in *trans*1/*cis*2). Thus

$$J_3(x,y,L_{23}\text{-},t) = v_{z3}\, G_3(x,y,L_{23}\text{-},t) - D\, \partial G_3(x,y,L_{23}\text{-},t)/\partial z \neq 0 \tag{33}$$

and

$$J_4(x,y,L_{23}\text{+},t) = v_{z4}\, G_4(x,y, L_{23}\text{+},t) - D\, \partial G_4(x,y,L_{23}\text{+},t)/\partial z \neq 0 \tag{34}$$

Continuity requires

$$J_3(x,y,L_{23}\text{-},t) = J_4(x,y,L_{23}\text{+},t) \tag{35}$$

In order for the particle to translocate successfully through DNP in the z direction so that it can be detected inside DNP, the net probability current at $L_{23}$ must be in the positive z direction. This can be achieved with a sufficiently large $V_{05}$. Thus

$$J_{34}(x,y,L_{23},t) = J_3(x,y,L_{23}\text{-},t) = J_4(x,y,L_{23}\text{+},t) > 0 \tag{36}$$

The behavior at the interface between DNP and *trans*2 is similar.

The tapered geometry of *trans*1/*cis*2 in Figure 4 aids drift of the particle into DNP. It can be modeled with a Fokker-Planck equation just as in Equation 12 but with a trapezoidal frustum boundary. The resulting system of equations is not as easily solved as Equations 12 through 18 although it is amenable to numerical solution. One obvious result is that the translocation time is decreased. Similar to the taper in *trans*1/*cis*2 aiding capture of the base at the entrance of DNP the abrupt increase in diameter from DNP to *trans*2 decreases the probability of a detected particle regressing into DNP from *trans*2. One can also think of these two behaviors in terms of entropy barriers [3]: the taper in *trans*1/*cis*2 decreases the barrier for entry into DNP (below what it would be with a rectangular box), while the step change going from DNP to *trans*2 effectively increases the barrier for a base regressing into DNP. Also as noted in Section III an extra drift potential across *trans*2 can provide additional inducement for the detected particle to pass through *trans*2 where it is recovered.

**Slowing down translocation**

The translocation of a base through DNP is too fast for the detector electronics. Methods to slow down translocation include the use of 'molecular brakes', magnetic or optical tweezers, alternative electrolytes such as LiCl, and increased salt concentration, see [3] for an overview. Here an approach based on the use of an electric field is proposed.

Let the electric fields over the five sections of the tandem cell be $E_{01}$, $E_{12}$, $E_{23}$, $E_{34}$, and $E_{45}$. Consider DNP in isolation. With a negative electric field $E_{34}$ over DNP of length $L_{34}$ = 10 nm, $D = 3 \times 10^{-10}$ m$^2$/sec, and $\mu = 2.4 \times 10^{-8}$ volt/m$^2$ sec, the data in Figure 5 show an increase in the mean translocation time, which indicates slowdown, but it is also accompanied by a significant increase in the variance. With $V_{34}$ approaching -0.25 V, the mean has increased by 7 orders of magnitude over the mean for $V_{34}$ = 0.25 V and the standard deviation is closely tracking the mean, indicating that diffusion has started to take over.

For this approach to work: 1) A cleaved base entering DNP must not regress into *trans*1/*cis*2; 2) A detected base exiting into *trans*2 must not regress into DNP; 3) The probability that there is more than one base in DNP must approach 0. To satisfy condition 1 a base moving from *trans*1/*cis*2 into DNP has to experience a positive drift field at the interface. This requires that $E_{23}$ and $E_{34}$ both be positive. To satisfy condition 2 a base moving from DNP into *trans*2 has to experience a positive drift field at the interface. This requires that $E_{34}$ and $E_{45}$ both be positive. Slowing down the base inside DNP requires $E_{34}$ to be negative. All three field sign conditions may be satisfied if $L_{34}$ is split into three parts $L_{34\text{-}0\ 34\text{-}1}$, $L_{34\text{-}1\ 34\text{-}2}$, and $L_{34\text{-}2\ 34\text{-}3}$ with respective electric fields $E_{34\text{-}0\ 34\text{-}1}$, $E_{34\text{-}1\ 34\text{-}2}$, and $E_{34\text{-}2\ 34\text{-}3}$ such that $E_{34\text{-}0\ 34\text{-}1} > 0$, $E_{34\text{-}1\ 34\text{-}2} < 0$, and $E_{34\text{-}2\ 34\text{-}3} > 0$. Such an electric field profile is shown in Figure 7.

The earlier analysis of DNP may be extended to the behavior of a base that experiences this kind of profiled potential in DNP. There is a tradeoff among the need to reduce the translocation speed through DNP, the need to prevent regression from DNP into *trans*1/*cis*2, and the need to prevent regression into DNP from *trans*2. Let $L_{34\text{-}0\ 34\text{-}1}$ = a $L_{34}$, $L_{34\text{-}1\ 34\text{-}2}$ = b$L_{34}$, and $L_{34\text{-}2\ 34\text{-}3}$ = c$L_{34}$, with a + b + c = 1. The first and second conditions require $E_{34\text{-}0\ 34\text{-}1}$ and $E_{34\text{-}2\ 34\text{-}3}$ both to be sufficiently positive. Since two electrodes are required to define the internal negative potential segment, each of a, b and c has a minimum value given by $a_{min} = b_{min} = c_{min} = e_w + e_s$, where $e_w$ = width of electrode and $e_s$ = interelectrode spacing. This spacing along with the applied voltages $V_{34\text{-}1}$ and $V_{34\text{-}2}$ can be used to determine the span of the negative electric field over DNP (Figure 7). (The voltages themselves need not be negative, it is the potential difference, and hence the corresponding electric field, that has to be negative.)

With this modification DNP can be represented as [Si pore-electrode-Si pore-electrode-Si pore]. It may be possible to achieve the desired field profile if ultra thin graphene sheets (which have been studied for their potential use in strand sequencing) are used as electrodes [13]. Figure 8 shows a schematic of the required modification, where voltages are applied to electrodes $V_0$, $V_{3\text{-}1}$, $V_{3\text{-}2}$, and $V_4$. Thus 49% of the potential difference $V_{0\ 3\text{-}1}$ drops across each of UNP and the segment $L_{34\text{-}0\ 34\text{-}1}$ and 0.5% across each of *cis*1 and *trans*1/*cis*2. A negative electric field exists across the segment $L_{34\text{-}1\ 34\text{-}2}$ with $V_{3\text{-}1} > V_{3\text{-}2}$. With $V_4 > V_{3\text{-}2}$, 99% of the potential difference $V_{3\text{-}2\ 4}$ drops across the segment $L_{34\text{-}2\ 34\text{-}3}$ and 1% across *trans*2.

The optimum electric field profile over DNP can be obtained by experiment. Here an estimate is obtained by using Equations 8 and 10 from the one-dimensional problem and ignoring the transitional behavior at the two ends. Let $V_{34\text{-}0\ 34\text{-}1} = V_{34\text{-}1} - V_{34\text{-}0} = V_a$, $V_{34\text{-}1\ 34\text{-}2} = V_b$, and $V_{34\text{-}2\ 34\text{-}3} = V_c$. With $V_a = V_c = 0.1$ volt and $V_b = -0.2$ volt the mean and standard deviation of the translocation time over each of the three segments of a nanopore of length $L_{34}$ = 10 nm are shown in Table 1 for different values of a and b. The translocation over the segment [$aL_{34}$, $aL_{34}$ + $bL_{34}$] is seen to be considerably slowed down by the negative field, which also dominates the total translocation time over DNP.

**Table 1**
**Translocation times over positive and negative electric field segments of DNP**

| a = c (positive field segment) | Mean ($10^{-8}$ sec) | Standard deviation ($10^{-8}$ sec) | b (negative field segment) | Mean ($10^{-3}$ sec) | Standard deviation ($10^{-3}$ sec) |
|---|---|---|---|---|---|
| 0.1 | 0.0365 | 0.0173 | 0.8 | 7.405078 | 7.405067 |
| 0.2 | 0.1458 | 0.0691 | 0.6 | 4.165356 | 4.165350 |
| 0.3 | 0.3281 | 0.1556 | 0.4 | 1.851269 | 1.851267 |
| 0.4 | 0.5834 | 0.2765 | 0.2 | 0.4628174 | 0.4628167 |

**Probability of bases arriving at DNP out of order**

Assume that bases are cleaved at a rate of one every T seconds. Without loss of generality let base 1 be cleaved at time t=0 and base 2 at t=T. Let $T_i$ = time for base i to diffuse-drift over *trans*1/*cis*2 and P = translocation time through the pore. $T_i$ and P are independent random variables with pdfs $f_{Ti}(.)$ and $f_P(.)$, mean $\mu_{Ti}=\mu_B$ and $\mu_P$, and standard deviation $\sigma_{Ti}=\sigma_B$ and $\sigma_P$ respectively, where the $T_i$s are assumed to be i.i.d. If the two pdfs are assumed to be unimodal [7, Figure 3] with finite support equal to six-σ, the supports would respectively be [max (0, $\mu_{Ti}-3\sigma_{Ti}$), $\mu_{Ti}+3\sigma_{Ti}$] and [max(0, $\mu_P-3\sigma_P$), $\mu_P+3\sigma_P$]. The following sufficient condition holds for the two bases to arrive in order:

$$T > (\mu_{T1}+3\sigma_{T1}) - \max(0, \mu_{T2}-3\sigma_{T2}) \tag{37}$$

In [17] the turnover rate for exonuclease under normal conditions is 10 to 80 msecs. Setting T=10 msecs and $V_{23}$=1.5 mV and interpolating over the data in Figure 6 gives $\mu_{T1}=\mu_{T2}$=1.6 msecs, $\sigma_{T1}=\sigma_{T2}$= 1.3 msecs. Using this in the inequality in Equation 37 gives

$$10 > 1.6 + 3 \times 1.3 - \max(0, 1.6 - 3 \times 1.3) = 5.4$$

This means that the bases arrive sequentially at DNP. Using similar arguments, it can be shown that detected bases passing into and through *trans*2 do so in their natural order.

**Probability of two bases being inside DNP at the same time**

Let bases be cleaved every T seconds. For bases 1 and 2 to be inside the pore at the same time base 2 must arrive in the pore before base 1 exits the pore. The condition for two bases not being in the pore at the same time is then obtained as

$$T + \max(0,\mu_{T2}-3\sigma_{T2}) > \mu_{T1}+3\sigma_{T1} + \mu_P+3\sigma_P \tag{38}$$

With T=10 msecs, $V_{34}$=200 mV, $V_{23}$= 1.6 mV, $\mu_{T1}=\mu_{T2}$=1.6 msecs, $\sigma_{T1}=\sigma_{T2}$= 1.3 msecs, $\mu_P$=0.000013 msec, and $\sigma_P$=0.000004 msec. Using these numbers in Equation 38

$$10 + \max(0, 1.6 - 3 \times 1.3) = 10 > 1.6 + 3 \times 1.3 + 0.000013 + 3 \times 0.000004$$

The two bases cannot be in the pore at the same time. The bandwidth required would be on the order of 4 Mhz.

Conversely, the minimum interval required between the release of two successive cleaved bases on the exit side of UNP so that they do not occupy DNP at the same time is given by

$$T_{min} = 3\sigma_{T1} + 3\sigma_{T2} + \mu_P + 3\sigma_P \qquad (39)$$

Using $\sigma_{T1} = \sigma_{T2}$= 1.6 msecs, $\mu_P$ = 0.000013 msec, and $\sigma_P$ = 0.000004 msec

$$T_{min} = 3 \times 1.6 + 3 \times 1.6 + 0.000013 + 3 \times 0.000004 = 9.600025 \text{ msecs}$$

In [5], T is given to be 1 to 10 msecs (compare with 10 to 80 msecs in [17]). With a suitable set of controls (temperature, salt concentration, etc.) the enzyme can be set to cleave bases at a rate that is < 1 every, say, 20 msecs.

With a negative electric field over part of DNP a rough bound can be obtained for the probability of two bases being in DNP at the same time. Consider for example $L_{34}$ = 10 nm, b = 0.4, $V_b$ = -0.15 volt, and $V_{23}$ = 1.6 mV. From the data in Figures 5 and 6 and Table 1, $\sigma_{T1} = \sigma_{T2}$ = 1.6 msecs, $\mu_P \approx 0.38$ msec, and $\sigma_P \approx 0.38$ msec. Using Equation 39

$$T_{min} = 3 \times 1.6 + 3 \times 1.6 + 0.38 + 3 \times 0.38 = 11.2 \text{ msecs}$$

This is within the range of turnover rates achieved with exonuclease as given in [17]. With a mean translocation time of 0.38 msec through a distance of $L_{34\text{-}2} = bL_{34} = 4$ nm, the detector bandwidth required would be on the order of 5 Khz, which comes close to the 1 msec translocation time criterion [4] for the effective detection of a base inside a nanopore.

## V  IMPLEMENTATION ISSUES
The tandem cell assumes a hybrid implementation with a biological nanopore for UNP and a synthetic one for DNP. A number of implementation issues are considered below.

**Positioning the enzyme**
The enzyme on the exit side of UNP need only be engineered so that it is in the path of the threading DNA sequence such that the first base of the remaining sequence is presented to it. If the leading base is not cleaved then it would mean that the ssDNA has either stopped moving or has slipped past the enzyme without being cleaved at all (since cleaving can only occur of the leading base). Failure to detect the characteristic pulses for the four (or more, if modified bases are considered) base types would indicate that no cleaving has occurred.

**Voltage drift**
With an ion-selective DNP pore, ion currents, which are typically on the order of a few 100 pA, can lead to an electro-osmotic potential which with the passage of time can cause buildup of charge in the pore and lead to the pore voltage drifting over time. Methods commonly used in electronic measurements may be used to solve the voltage drift problem. One of these is to use a stable reference voltage against which the drift is tracked and the difference subtracted from the recorded data (similar to the moving average in statistical trend analysis of time series data). Alternatively the *trans*1/*cis*2 and *trans*2 compartments and DNP can be drained periodically and refilled with electrolyte. To prevent the occurrence of deletion errors due to cleaved bases still in transit through *trans*1/*cis*2 while draining is taking place, the draining step may be preceded by retraction of the strand in UNP (achieved by temporarily lowering or reversing the potential $V_{05}$) and pausing until the cleaved bases in transit have passed into DNP and been detected through their characteristic blockade levels.

**Cleaved bases sticking to the side wall of  *trans*1/*cis*2**
A cleaved base may stick to a side wall while diffusing inside *trans*1/*cis*2. The probability of this event can be calculated using the model in Section IV. Bases can be prevented from sticking by holding the side wall of *trans*1/*cis*2 at a slightly negative potential (the  appropriate value to use can be determined experimentally) with respect to $V_1$. This effectively creates a reflecting wall for the negatively charged base. Alternatively the compartment walls may be chemically treated to prevent sticking. For example, the use of a lipid coat to control clogging of a translocating protein in a solid-state nanopore is described in [20].

**Solid state pore for DNP**
A solid state pore has the advantages of scaling and integration in fabrication and has been studied widely both experimentally and theoretically in the context of DNA sequencing. While such pores are not useful in strand sequencing for single-nucleotide discrimination because of their thickness (currently they have a minimum thickness of 20 nm and an hourglass shape with actual pore thickness of 10 nm [10]), with exonuclease sequencing using a tandem cell this may not be a problem because of the near zero probability of two nucleotides being in the nanopore at the same time, as shown above. They are also easier to fabricate because of the relaxed tolerances on both electrode width and electrode spacing, especially an advantage if a negative electric field over DNP is used to slow down translocation.

**Negative field over DNP, graphene electrodes, graphene pores**
Implementing a negative field over part of DNP could take the form of two graphene sheets interposed between three layers of silicon pores and acting as electrodes across which the negative field is applied. The resulting DNP would then have the structure [Si pore-graphene electrode-Si pore-graphene electrode-Si pore]. See Figure 8.

   A graphene nanopore by itself can be used for exonuclease sequencing with a tandem cell. The cell would then have the structure [*cis*1, UNP, *trans*1=*cis*2, DNP, *trans*2], where DNP is [Si pore-graphene electrode-Si pore-graphene sheet-Si pore-graphene electrode-Si pore]. The translocating cleaved base is then identified by the level of the transverse current through the graphene sheet [13].

   Negative fields may also be useful for translocation slowdown in strand sequencing, for which nanopores in graphene nanoribbons have been investigated recently [11-13]. The translocation speed can be controlled by embedding the graphene layer in the negative-field segment of a solid-state nanopore with an electric field profile similar to that in Figure 7. A cell would then have the structure [*cis*-Si pore-graphene electrode-Si pore-graphene sheet-Si pore-graphene electrode-Si pore-*trans*], where the large thickness of the Si pore, by itself a disadvantage in sequencing, is no longer important.

In both strand and exonuclease sequencing based on the above approach, the µA-level transverse currents through the graphene sheet combined with the negative-field-based slowdown of strand or cleaved base could provide an efficient sequencing method with a lower detection bandwidth and S/N ratios that are significantly higher than with a method that relies on discrimination based only on axial ionic pore currents of ~100 pA.

## VI DISCUSSION

1. With a tandem cell, repeat bases (homopolymers) do not present a problem because of the time and space separation between successive cleaved bases.
2. An arbitrary number of tandem cells could be implemented in parallel. With a sequencing rate of 100/second, an array of 10000 cells can potentially sequence a billion ($10^9$) bases in ~16 minutes.
3. A pipeline of tandem cells with a *cis trans-cis ... trans-cis trans* structure may be used for error checking and/or to obtain upstream-downstream correlations in real time. With an N-stage pipeline N× coverage is possible, requiring, at least in principle, little more time than the time needed for 1× sequencing.
4. A recent report describes the use of heavy tags attached to mononucleotides that are used by a processive enzyme to synthesize DNA threaded through the enzyme [21]. When a nucleotide is added to the growing DNA strand, the tag is cleaved and drops through a nanopore causing a blockade event that is unique for each of four different tag types. If the tags could be attached to bases in ssDNA, this method could be adapted for use with the tandem-cell method proposed here. The heavier base-specific tags could lead to better discrimination among the base types.
5. A modified version of the tandem cell may be considered for polypeptide sequencing. Such a structure would be more complicated than the one for DNA sequencing because: a) peptides can be positively or negatively charged or charge neutral; and b) different peptidases have to be used for different types of peptides. This will require cycling the shrinking polypeptide through a pipeline of tandem cells one for each type of peptide as well as an algorithm to assemble the sequence from the signal data consisting of multiple time series, one for each peptide type.

# Figure Captions and Figures

**Figure 1. Schematic of nanopore DNA sequencing: (a) strand sequencing (b) exonuclease sequencing.** AHL nanopore ~10 nm long has broad vestibule (length 5 nm) and narrow barrel/stem (length 5 nm, smallest diameter 2 nm). Pore is embedded in lipid bilayer (membrane) ~5 nm thick.

**Figure 2. Schematic of proposed tandem cell.** (a) Tandem cell with five sections or stages in pipeline: *cis*1, UNP (depicted here as an AHL pore), *trans*1 (= *cis*2), DNP (depicted here as a solid-state pore), and *trans*2. Dimensions considered in the text are: 1) *cis*1: box of height 1 μm and cross section 1 μm$^2$; 2) UNP: AHL pore of length 8 nm and diameter 2 nm; 3) *trans*1/*cis*2: tapered box of length 1 μm tapering from cross-section of 1 μm$^2$ to 4 nm$^2$; 4) DNP: solid-state pore of length 10-20 nm and diameter 2 nm; 5) *trans*2: box of height 1 μm and cross-section 1 μm$^2$. (b) Voltage profile. Electrodes assumed inserted at top of *cis*1 and *trans*2. Negative field across graphene electrodes inserted laterally into DNP slows down translocating base (see text and Figure 8).

**Figure 3. Tandem cell with reinforcing potential difference in *trans*2.** Two electrodes in *trans*2, instead of one; used to apply reinforcing drift field to prevent detected base from regressing into DNP.

**Figure 4. Coordinate systems for models.** Different coordinate systems used for (a) Stage 4, (b) Stage 3, and (c) Stage 3. Dimensions used: (a) L= 8-10 nm; (b) L = 1 μm, $w_p$ = 1 μm; (c) $L_t$ = 1 μm, $w_t$ = 1 μm, $w_p$ = 2 nm.

**Figure 5. Translocation statistics of DNP.** Mean and standard deviation of time for particle to translocate from time of entry into DNP (negligible cross-section and length L = 8-10 nm) to time of exit into *trans*2. Parameter values used: mononucleotide mobility μ = 2.4 × 10$^{-8}$ m$^2$/volt-sec, diffusion constant D = 3 × 10$^{-10}$ m$^2$/sec. Calculations are for typical absolute potential difference in the range 0.1-0.3 volt [5]. Negative field across DNP results in markedly decreased translocation times, sufficient to reduce detection bandwidth significantly.

**Figure 6. Translocation statistics of *trans*1/*cis*2.** Mean and standard deviation of translocation time for particle (cleaved base) released by exonuclease at top of *trans*1/*cis*2 (= 3-dimensional box with height 1 μm and cross-section 1 μm$^2$) to move to entrance of DNP. Parameter values used: mononucleotide mobility μ = 2.4 × 10$^{-8}$ m$^2$/volt-sec, diffusion constant D = 3 × 10$^{-10}$ m$^2$/sec. Calculations for cell voltages of 0.1-0.3 volt, with ~1-2 mV dropping across *trans*1/*cis*2.

**Figure 7 Voltage and electric field profiles over DNP.** Example of a profiled voltage in which the pore length is divided into three segments $L_{34-1}$, $L_{34-2}$, and $L_{34-3}$, with lengths $L_{34-1} + L_{34-2} + L_{34-3}$ = $L_{34}$. Electric field is positive over $L_{34-1}$ and $L_{34-3}$, negative over $L_{34-2}$. (The voltages themselves need not be negative. Thus $V_{34-1} - V_{34-0} > 0$, $V_{34-3} - V_{34-2} > 0$, and $V_{34-2} - V_{34-1} < 0$.) Also $V_{34-0} > V_{23}$ (voltage across *trans*1/*cis*2) and $V_{34-3} < V_{45}$ (voltage across *trans*2).

**Figure 8. Tandem cell modified for negative field over segment of DNP.** Graphene electrodes laterally inserted into DNP. Negative field applied over middle segment of DNP

through graphene electrodes set to voltages $V_{3\text{-}1}$ and $V_{3\text{-}2}$. Thus $V_0 < V_{3\text{-}1}$, $V_{3\text{-}2} < V_{3\text{-}1}$, $V_{3\text{-}2} < V_5$, where $V_0$ and $V_5$ represent electrodes in *cis*1 and *trans*2

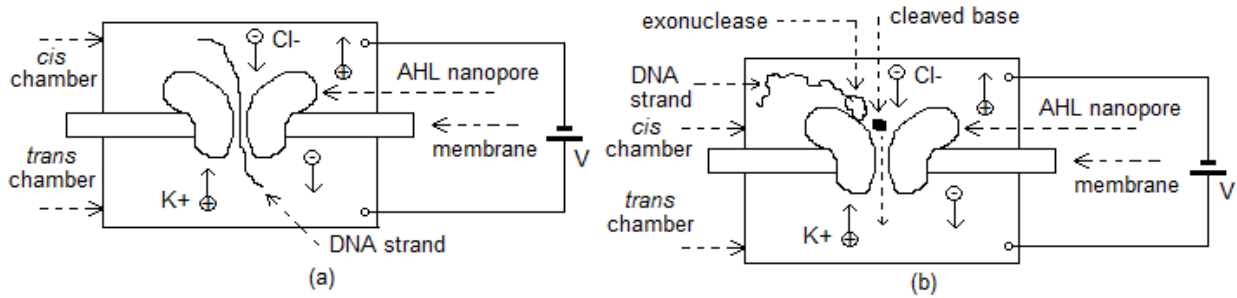

**Figure 1**

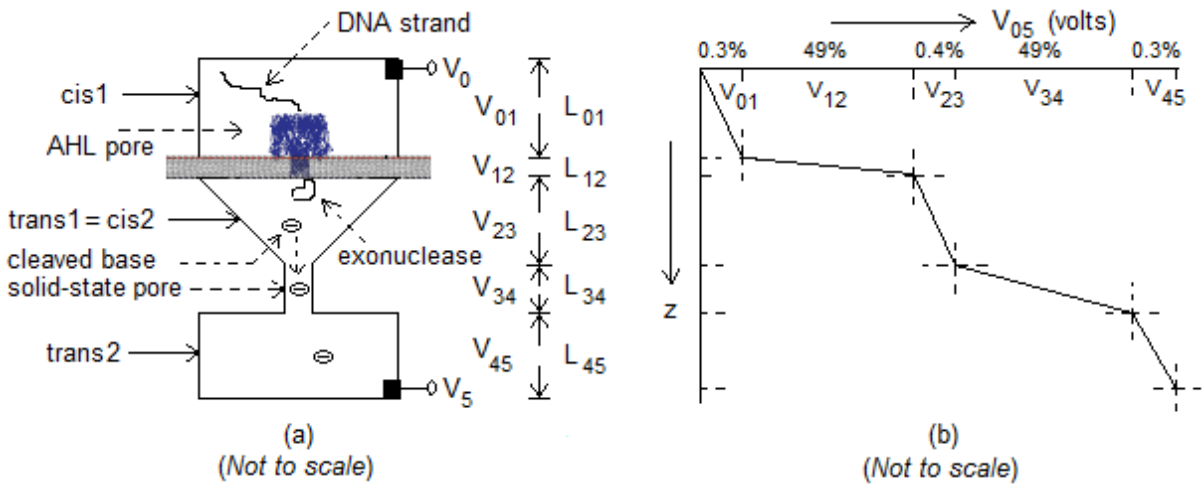

**Figure 2**

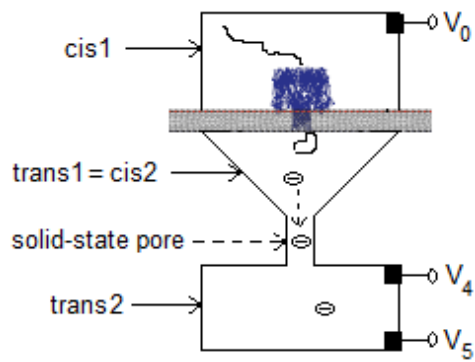

**Figure 3**

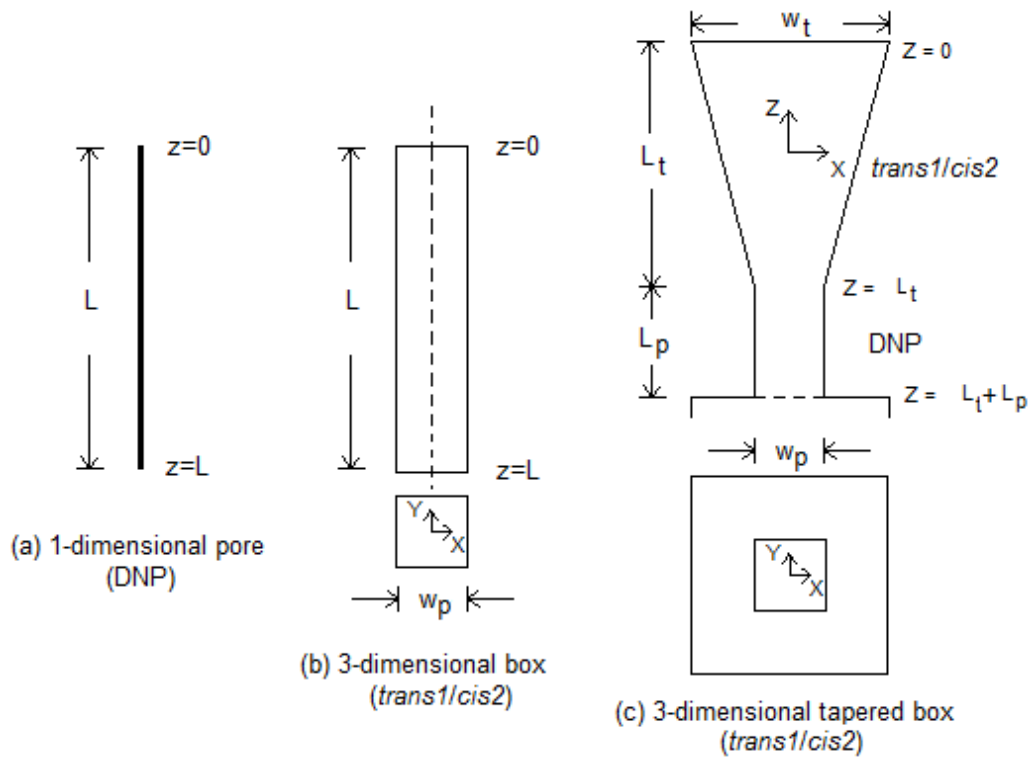

**Figure 4**

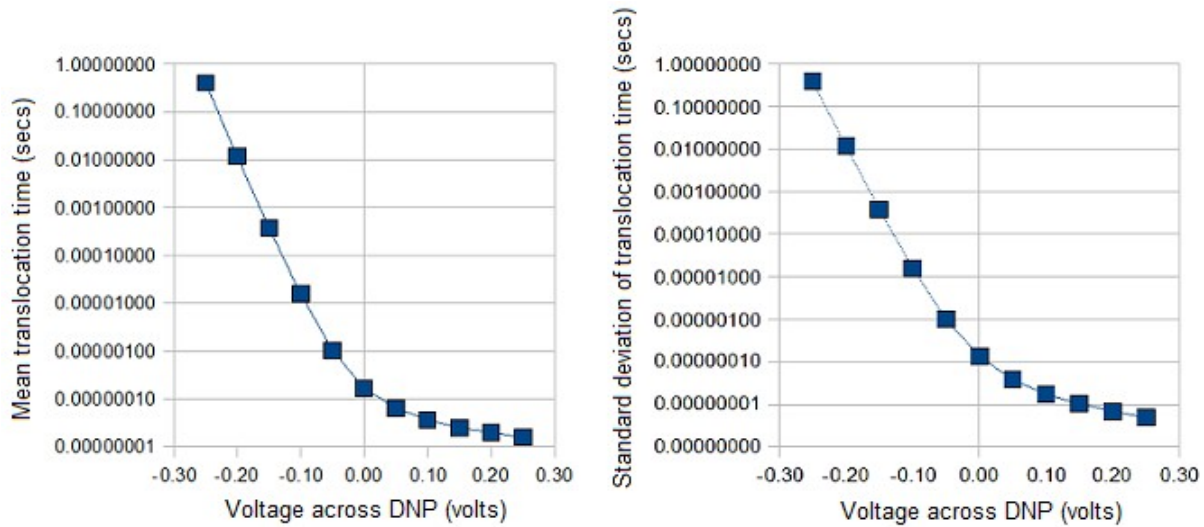

**Figure 5**

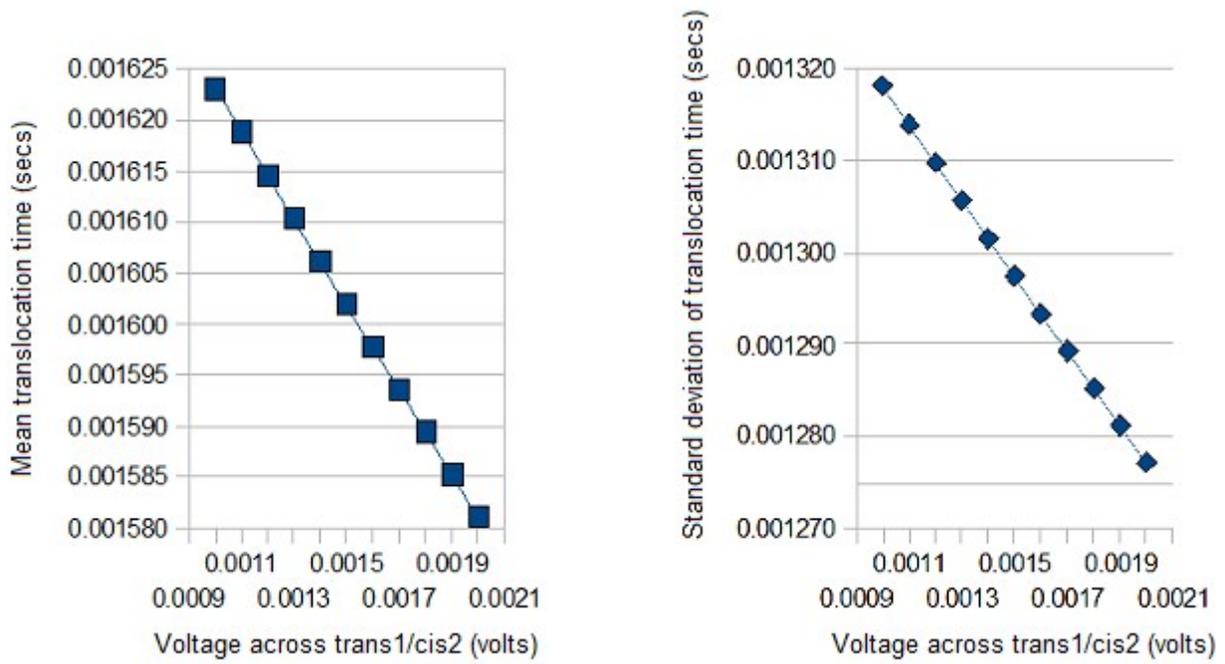

**Figure 6**

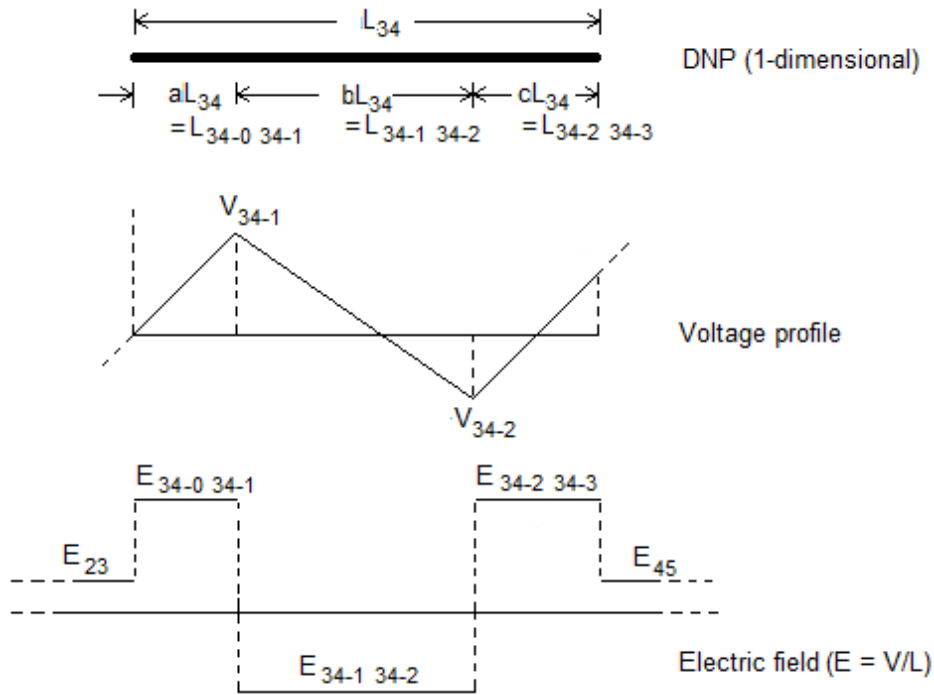

**Figure 7**

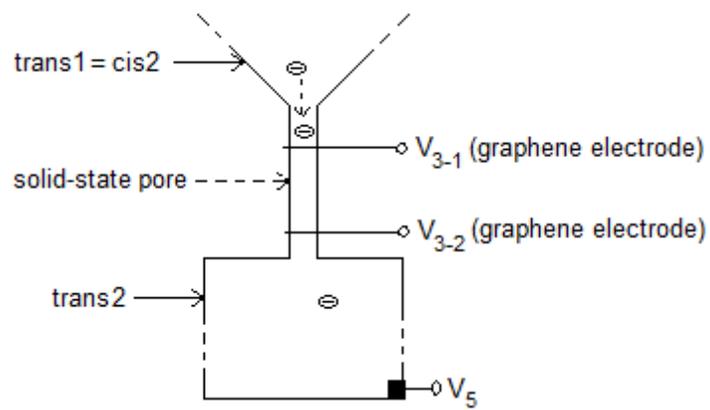

**Figure 8**